**On measuring the acoustic state changes in lipid membranes using fluorescent probes**

*Shamit Shrivastava[1]\*, Robin O. Cleveland[1], Matthias F. Schneider[2]*

1. *Institute for Biomedical Engineering, University of Oxford, Oxford OX3 7DQ, UK*
2. *Medizinische und Biologische Physik, Technische Universität Dortmund, Otto-Hahn Str. 4, 44227 Dortmund, Germany*

*\*Corresponding Author: shamit.shrivastava@eng.ox.ac.uk*

*Draft 6 November 2018*

Ultrasound is increasingly being used to modulate the properties of biological membranes for applications in drug delivery and neuromodulation. While various studies have investigated the mechanical aspect of the interaction such as acoustic absorption and membrane deformation, it is not clear how these effects transduce into biological functions, for example, changes in the permeability or the enzymatic activity of the membrane. A critical aspect of the activity of an enzyme is the thermal fluctuations of its solvation or hydration shell. Thermal fluctuations are also known to be directly related to membrane permeability. Here solvation shell changes of lipid membranes subject to an acoustic impulse were investigated using a fluorescence probe, Laurdan. Laurdan was embedded in multi-lamellar lipid vesicles in water, which were exposed to broadband pressure impulses of the order of 1MPa peak amplitude and 10μs pulse duration. An instrument was developed to monitor changes in the emission spectrum of the dye at two wavelengths with sub-microsecond temporal resolution. The experiments show that changes in the emission spectrum, and hence the fluctuations of the solvation shell, are related to the changes in the thermodynamic state of the membrane and correlated with the compression and rarefaction of the incident sound wave. The results suggest that acoustic fields affect the state of a lipid membrane and therefore can potentially modulate the kinetics of channels and proteins embedded in the membrane.

**Introduction**

The outer membrane of biological cells is the primary interface through which they interact with the rest of the world. These membranes are self-assembled structures mainly formed of lipids and proteins, that play a fundamental role in controlling the flux of material across the boundaries of a cell. The self-assembly process is driven by the thermodynamic potential of the system and its curvature is determined by the thermodynamic susceptibilities of the system, such as isobaric specific heat $C_p$ and adiabatic compressibility $\kappa_s$, which can be measured experimentally. Therefore, the relaxation of any out of equilibrium perturbation of the membrane is closely related to the thermodynamic susceptibilities of the system (1). These susceptibilities are typically functions of pressure and temperature for a system of given composition, but also depend on other state variables such as pH and membrane potential. The acoustic absorption of lipid membranes, and indeed any medium, is dependent on $C_p$ and $\kappa_s$ and this has been shown independent of membrane composition (2, 3). Therefore, acoustic absorption will also depend on other state variables, for example, acoustic absorption studies of an aqueous suspension of globular proteins showed a marked pH dependence and a peak near conformational transitions, such as fluid-gel transitions in lipids and pKa of proteins (4, 5). This absorbed acoustic energy has the potential to be transduced to a function of embedded molecules, i.e. to a change in enzymatic or channel activity. The question is of particular interest for applications of acoustics in non-invasive therapies, such as sonoporation or neuromodulation(6). Here we directly measure the acoustic modulation of the emission spectrum of fluorescent molecules embedded in lipid membranes.

Fluorescence phenomenon has a long history in exposing molecular mechanisms during fast processes in biological membranes. One example is a nerve impulse, where it has been shown that based on emission spectra of 8-Anilino-1-naphthalene sulfonic acid (ANS) that the nerve impulse is accompanied by conformational changes in membrane macromolecules(7, 8). Since then, many such fluorescent probes known as "solvatochromic" dyes, such as Laurdan, Prodan, and di-4-aneppdhq (9–23) as well as "electrochromic" (voltage sensitive) dyes such as di-8-anneps and RH421(24), have been developed. These dyes are known to have spectral properties that are sensitive to their solvation or the dielectric environment. Fluctuations in the solvation shell have been shown to affect the enzyme activity and have also been shown to be correlated with changes in the emission spectrum of dyes. In particular, the fluctuations in enthalpy H, volume $V$, and dipole moment $M$ of the solvation shell, scale as;

$$< (\Delta H)^2 > \sim C_p T^2, < (\Delta V)^2 > \sim k_T V T \ and \ < (\Delta M)^2 > \sim T V \varepsilon;$$

where $C_p$, is bulk heat capacity, $k_T$ isothermal compressibility, $\varepsilon$ dielectric susceptibility and $T$ temperature (25–27). Laurdan, which shows a redshift as a membrane becomes more hydrated, has been used extensively in membrane studies and was selected for this study. The goal here is to use changes in the fluorescence emission of Laurdan in response to acoustic stimulation to provide insights into the mechanism of mechanotransduction of acoustic radiation.

## Material and Methods

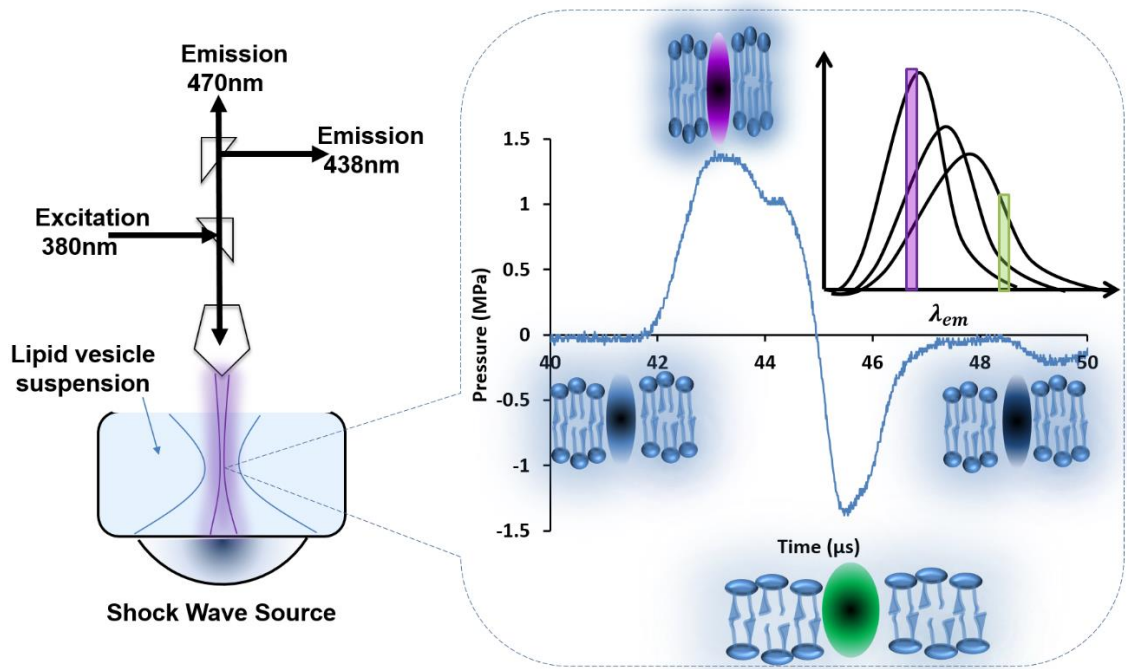

**Figure 1 (a) Experimental Setup** *A shock wave source coupled to a water tank provides a pressure impulse at its focus which coincides with the focus of the optical setup. (b) An acoustic waveform measured at the focus for energy level 5 exhibiting peak pressures of $\pm 1.30 MPa$ and a cartoon depicting the relationship between lipid structure and the shift in the wavelength of the optical emission. The shift is quantified ratiometrically by the two photomultiplier tubes recording simultaneously at two different wavelengths (438 and 470nm) as shown in (a).*

## Materials:

The lipids 1,2-dioleoyl-sn-glycero-3-phosphocholine (DOPC 850375), 1,2-dimyristoyl-sn-glycero-3-phosphocholine (DMPC 850345), 1,2-dipalmitoyl-sn-glycero-3-phosphocholine (DPPC 850355) were purchased as a 25 mg/mL solution in chloroform from Avanti Polar Lipids, Inc., (Alabaster, AL, USA). Laurdan was purchased from Life Technologies, (CITY, STATE, USA). Unless otherwise stated, all other chemicals were purchased from Sigma Aldrich (UK?).

**Preparation of multilamellar vesicles (MLVs):**

100 ul of the lipids (DOPC, DMPC or DPPC) (25mg/ml), from the chloroform stock without any further purification, were dried in a 10 ml glass vial overnight in a vacuum desiccator. The dried film was rehydrated in 5 ml MilliQ ultrapure water and heated for 30 mins on a hot plate at 50℃ with the cap tightly sealed. The film was then vortexed at 2000 rpm for 30 seconds and the processes repeated once more to get the MLVs suspension. Laurdan in $10\mu l$ DMSO (dimethyl sulfoxide) was added to 5ml of MLV suspension to get an approximately 1:100 dye to lipid ratio, which was then mixed with 500ml of additional MilliQ water in the temperature regulated tank.

**Optical and pressure measurements and data analysis:**

A custom optical setup was built for exposing a sample to shock waves and simultaneously measuring the fluorescence signals from the sample at two wavelengths (fig. 1). Shock waves were generated by a piezoelectric focused source (Swiss Piezoclast, EMS, Switzerland). The pressure waveform at the focus was measured using (i) a PVDF needle hydrophone (Dr. Muller Instruments, Oberursel, Germany) with a manufacturer-specified sensitivity of 12.5mV/bar (bandwidth $0.3 - 11$ MHz $\pm 3.0$ dB) and (ii) a PZT hydrophone (TC4013 Teledyne RESON, Denmark) (bandwidth 1Hz to 170kHz). The source was operated at energy setting 5 (unless otherwise stated) and shock waves delivered at 3 Hz. A representative waveform measured by the PVDF membrane hydrophone is shown in Fig 1(b). Temperatures were measured at the optical focus using a needle thermocouple (HH611A PL-4, Omega, UK), which was removed during shock wave exposure.

The fluorescence was excited by a high power light emitting diode (M385LP1, Thorlabs, UK) with a central wavelength of 385nm followed by a bandpass filter (390 ∓ 20nm). The LED was triggered 5 ms before every acoustic impulse for a duration 10 ms. A dichroic mirror (414 nm) directed the excitation light through an objective with a 13 mm working distance (LWD M PLAN FL 20X Olympus, Japan) into the sample and a region of $400\mu m$ in diameter was illuminated. The fluorescent light emitted by the sample focused at the back focal plane of the objective passed through the first dichroic mirror and was then split by :a second dichroic (452 nm). The light from the two beams was detected by two photomultiplier tubes (H10493-003, Hamamatsu, Japan) one with a 438 ($\mp$ 12nm) bandpass filters and the other with a 470 nm ($\mp$ 11nm) bandpass filter. The PMTs had a signal bandwidth of 8 MHz.

Data from the PMTs was acquired using NI PCI 5122 dual channel simultaneous 100Ms/s digitizer for 500 $\mu s$ of which 50 $\mu s$ preceded the trigger and was analyzed using NI Labview. A 200kHz digital low pass filter was applied (3$^{\text{rd}}$ order Butterworth). The first 10us were averaged to obtain the baseline intensity $I_0$ for each channel from which the relative intensity change was calculated: $\frac{\Delta I}{I_0} = \frac{I}{I_0} - 1$.

The shift in the emission spectrum was characterized by a normalized ratiometric parameter

$$\Delta RP / RP = \left(\frac{\Delta I}{I_0}\right)_{438nm} - \left(\frac{\Delta I}{I_0}\right)_{470nm}, \tag{1}$$

where $RP = I_{438nm} / I_{470nm}$. As shown in Appendix A the ratiometric parameter is related to the change in the wavelength of the maximum emission and for the case of Gaussian distribution of width $\Gamma$ the relationship is:

$$\Delta RP / RP = \frac{\Delta \lambda_{max}}{\Gamma^2} 2(\lambda_2 - \lambda_1) \tag{2}$$

The advantage of the ratiometric parameter is that it is robust to artifacts as opposed to RP alone, such as concentration inhomogeneities, background fluorescence and purely mechanical undulations in the field of view (e.g. motion of surfaces in the experimental setup) as these artifacts have the same impact on intensity at both wavelengths (see Appendix A). $\Delta RP / RP$ is invariant of the optical setup as well and represents the changes in the thermodynamic state of the membrane. In the discussion and Appendix B it is shown that changes in the maximum emissions wavelength can be related to changes in the enthalpy of the membrane, that is, $\Delta \lambda_{max} \sim \Delta H$. Therefore, changes in the ratiometric parameter allows us to detect changes in the thermodynamic state of the membrane independent of other factors.

**Results and Discussion**

The initial thermodynamic state of different MLV suspensions was quantified by measuring RP as a function of the reduced temperature (Fig 2). The reduced temperature represents the proximity of lipid vesicles to their phase transition temperature $T_{pt}$ (i.e., melting point) and is defined as $T^* =$ $\frac{T-T_{pt}}{T_{pt}}$ where absolute temperatures are employed. DPPC, with the typical main fluid-gel phase transition temperature $T_{pt} = 41.3^o$ C (28), has a T*=-0.067 at 20ºC and is therefore in the "coldest" state of the samples tested and the resulting ordered state of the lipids produces the highest RP=1.3. DMPC with a typical $T_{pt} = 23.6^o C$ (28, 29), was measured at $10^o C$, $23^o C$, and $37^o C$ and RP showed a monotonic decrease as the state became more disordered. DOPC with $T_{pt} = -20^o C$ (30) represents the most disordered at $20^o C$, and resulted in the smallest RP=0.61. This data supports the concept that the ratiometric parameter can be used to capture the thermodynamic state of lipids.

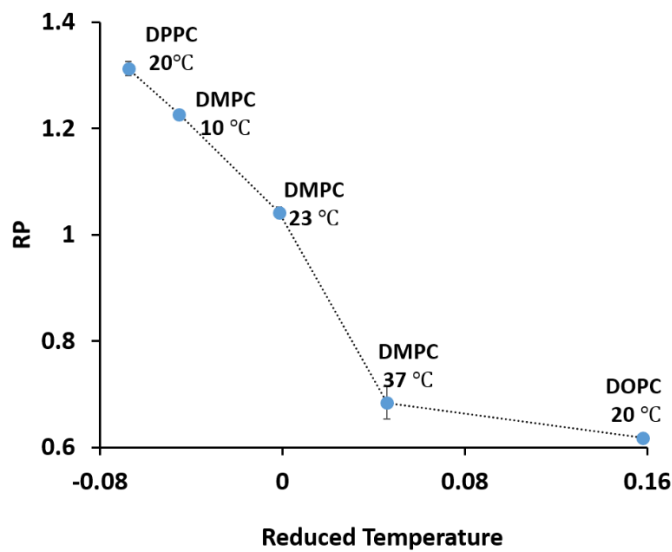

**Figure 2 Initial conditions** *The equilibrium RP value calculated from measured Laurdan emission in different MLV formulations, plotted as a function of the reduced temperature. There is a monotonic decrease in RP as the lipids go from an ordered or "solid" state (T\*<0) to a disorder or "fluid" state (T\*>0).*

Figure 3 shows $\frac{\Delta RP}{RP}$ as a function of time for all five lipid preparations in response to an acoustic impulse which was fired at 0 μs. The waveforms are averages over 500 measurements. For the two lipids with T*<0, there was no detectable response which is consistent with the membranes being in a gel state. For DMPC at 23℃ (T*~0), a relatively large negative spike $\frac{\Delta RP}{RP} \approx -0.004$ was detected at 45 μs (which corresponds to the propagation time from the shock source to the sample), which suggests strong coupling into the membrane when it is at the phase transition.   For the two lipids in the fluid state ($T^* > 0$) a response was detected when the shock wave arrived but it was smaller in amplitude( $\Delta RP/RP \approx -0.0015$) than the $T^* = 0$ case.

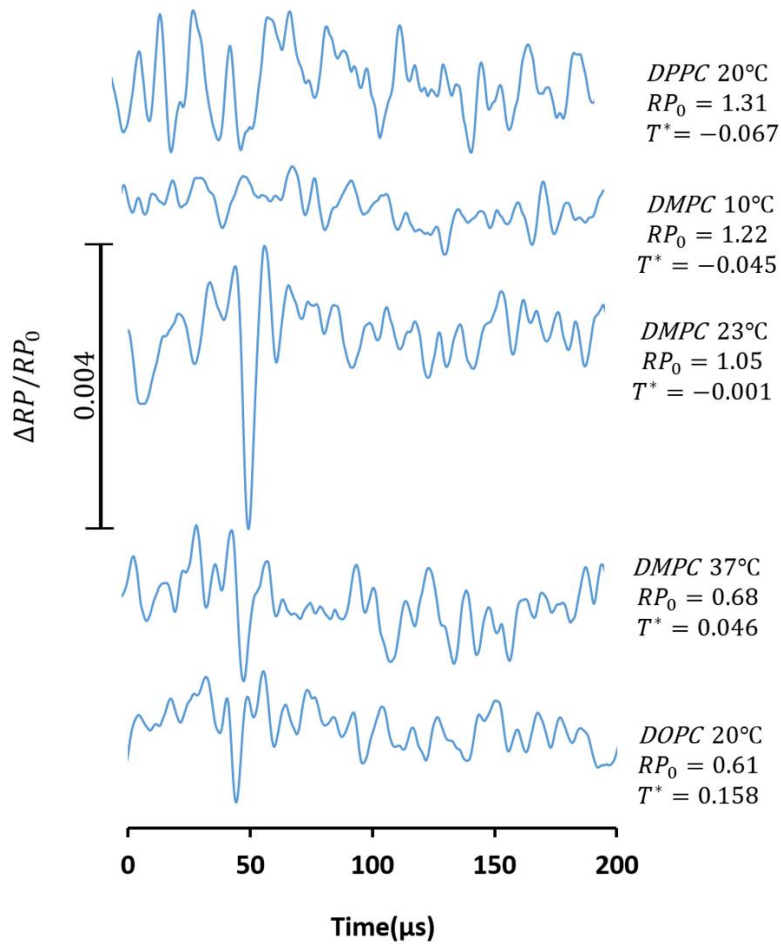

**Figure 3 Perturbation response as a function of membrane phase state** *The average responses (n=500, 200 kHz Low-Pass filter) of the vesicles subject to a shock wave are quantified by the change ΔRP/RP plotted as a function of time. The shock wave is triggered at t=0 and arrives at the sample at t=50 µs.  The maximum response occurred for DMPC at* 23℃ *which was closest to its transition temperature.  The scale bar on the y-axis corresponds to ΔRP/RP=0.004.*

Instead of calculating $T^*$ based on transition temperatures reported in the literature, the temperature corresponding to a maximum in $-\frac{dRP}{dT}$ can be used to estimate the transition temperature in these experiments. This is based on the empirical observation that response function or the susceptibility of a system to an external perturbation is maximum at the transition. Figure 4 shows this behavior for homogeneous DMPC as well as inhomogeneous vesicles composed of 50% DMPC and 50% DPPC by weight. Accordingly, a transition temperature of 23.45℃ was estimated for DMPC vesicles and 31.9℃ for 50:50 DMPC: DPPC (w/w).

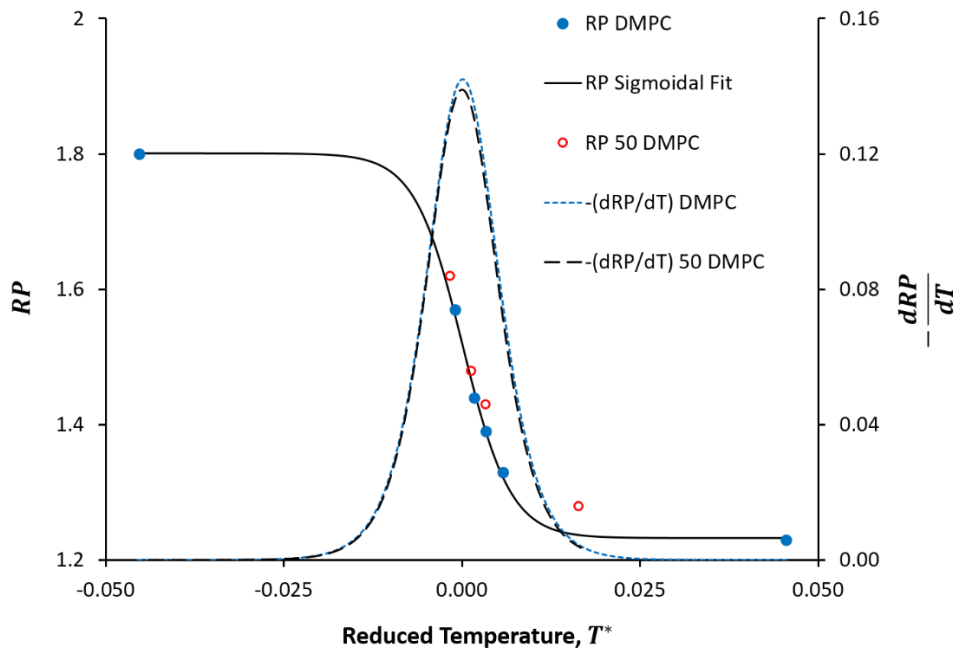

**Figure 4 Transition temperatures estimated from the temperature dependence of spectral shift.** *RP values measured at discrete temperatures in DMPC and DMPC: DPPC 50:50 (w/w) vesicles. Sigmoidal fits were produced for both vesicles to find the peak in the derivative $-\frac{dRP}{dT}$. The figure plots the fit only for the DMPC vesicles. The maximum in the derivative with respect to the temperature provides an estimate for the phase transition temperatures in the two systems. The two peaks line up at $T^* = 0$ for transitions temperatures of 23.45℃ for DMPC vesicles and 31.9℃ for 50:50 DMPC: DPPC (w/w), respectively.*

Figure 5 shows waveforms measured in DMPC for temperatures close to transition. Clearly, the maximum responses are observed in the vicinity of the transition, which was estimated to be at 23.45℃ (Fig.4). However, it can be seen that for 24℃ and 25.2℃ ($T^* > 0$) the maximum values of ${\Delta RP}/{RP}$ are similar in amplitude. This suggests an asymmetry around the phase transition, where the impulse response changes sharply as a function of temperature on the gel-side of the transition or $T^* < 0$, compared to the fluid-side or $T^* > 0$. This is consistent with the general behavior of the thermodynamics susceptibilities of lipid systems near the melting transition, where the rise in susceptibility is sharper on the gel side compared to the fluid side(2, 31).

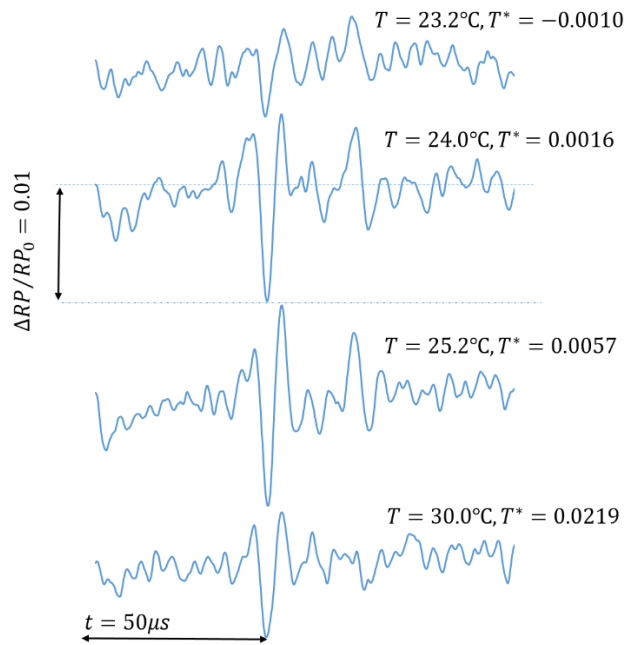

**Figure 5 Perturbation response as a function temperature near a phase transition** *Average $\Delta RP/RP$ waveforms (n=500, 200kHz low pass filter) for DMPC vesicles near the main transition. The scale bar on the y-axis corresponds to $\Delta RP/RP=0.01$.*

To demonstrate that $\Delta RP/RP$ can also be used to measure the macroscopic state of heterogeneous lipid membranes, vesicles were formed using a 1:1 (w/w) mixture of DMPC and DPPC which is reported to have a cooperative phase transition around 32℃ (32). Figure 6 shows measurements of the heterogeneous vesicles in response to shock waves at different temperatures. The greatest response was measured to be at 32.4℃ with a maximum $\Delta RP/RP \sim 0.019$ which is consistent with the estimated location of the phase transition at 31.9℃. Therefore, the variation in the amplitude of $\Delta RP/RP$ with membrane states indicates that Laurdan, and in particular the ratiometric parameter, can capture the thermodynamic state of the membrane and the maximum response is observed near a phase transition ($T^* = 0$). This is consistent with the previous observations in lipid suspensions where acoustic absorption was found to be maximum near phase transition(2) and hence transfer of energy from the shock wave to the lipid membrane should be greatest in this state.

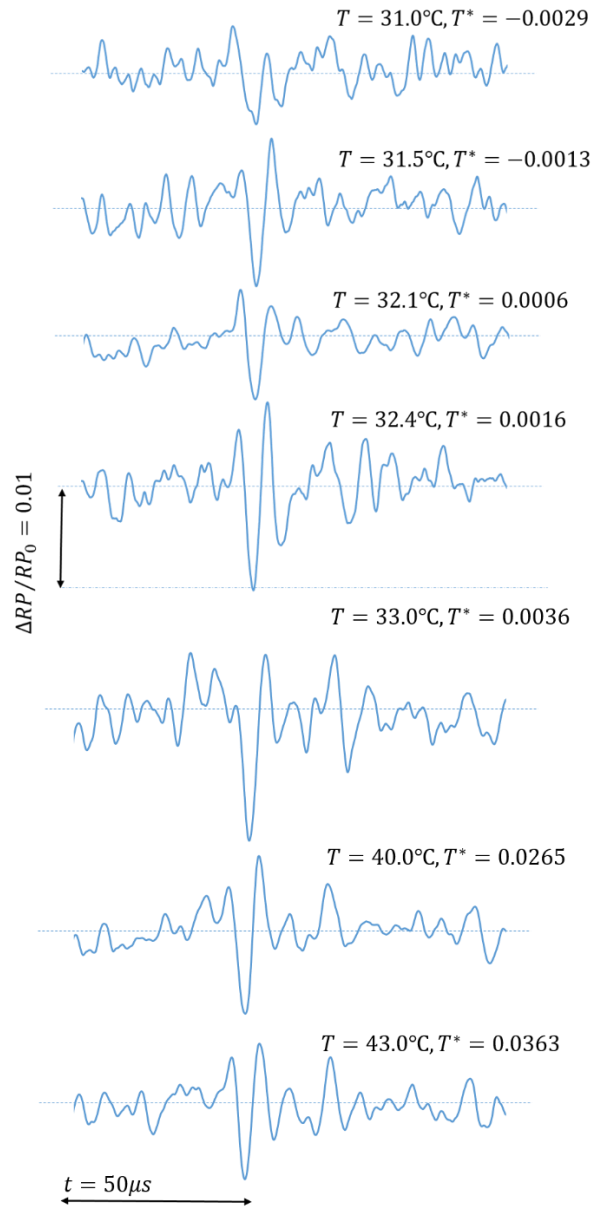

**Figure 6 Material invariance of response-state relationship** *The average responses of vesicles made of 50:50 (w/w) mixture of DMPC and DPPC near the corresponding transition* (32) *(n=500,*

*Low-Pass 200kHz) are quantified by ΔRP/RP, plotted as a function of time. The scale bar on the y-axis corresponds to ΔRP/RP=0.01.*

A first-order quantitative relationship between $\Delta RP/RP$ and the specific volume can be determined from the static measurements in Fig. 2. For DMPC the ratiometric parameter changes by -0.44 as DMPC goes from gel to fluid state. The increase in specific volume during the gel-fluid phase transitions under quasistatic conditions is reported to be $\frac{\Delta V}{V} \approx 0.03$ (3, 33). Therefore, the relationship between the ratiometric parameter and the specific volume is:

$$\Delta RP/RP = -14.6 . \Delta V/V$$

Thermodynamically, the relative fluctuations in volume $\frac{\Delta V}{V}$ can be related to acoustic pressure (in the linear approximation) as $k_s \Delta P = -\frac{\Delta V}{V}$, where $k_s$ is the adiabatic compressibility of the system. Therefore the ratiometric parameter can be related to acoustic pressure through $\Delta RP/RP = 14.6 k_s \Delta P$. The incident pressure is of the order 1.5 MPa and for DMPC in the gel state $k_s$=3.1 × $10^{-10}$ $m^2/N$ from which $\Delta RP/RP = 0.007$ whereas for the fluid state $k_s= 5.27 \times 10^{-10}$ $m^2/N$, and $\Delta RP/RP = 0.01$. These compare favorably to the observed maximum $\Delta RP/RP$ values of 0.015 when one considers that $k_s$ was measured at a fixed frequency of 5MHz and at low radiation intensity in a resonator cavity (2).

Figure 7 shows an overlay of the $\Delta RP/RP$ waveform (optical DC to 1MHz) for 50:50 DMPC: DPPC lipid and the pressure waveforms (mechanical) from the PVDF hydrophone (bandwidth 0.3 MHz to 11 MHz) and the PZT hydrophone (bandwidth 1 Hz to 170 kHz). It can be seen that the RP waveform closely resembles the lower bandwidth pressure waveform while the solvation relaxation time for Laurdan is of the order of nanoseconds, it senses these conformational changes via intermolecular interactions with relaxation timescales of the order of 1-10$\mu s$ in lipid membranes that corresponds to lateral diffusion(34).

From a mechanism perspective, we expect that the dye does not have a significant response of its own at the pressure amplitudes up to timescales corresponding to 1 MHz, and it only responds to the thermodynamic state of the collective system. The reason for the observed changes in RP is that as the thermodynamic state of the membrane changes, so does the energy distribution of the dye molecules, which results in the wavelength shift of the emission spectra. Understanding how this distribution is altered during a thermodynamic process is key to a mechanistic understanding of the observed change in the spectrum and its relation energy landscape of a dye+membrane system.

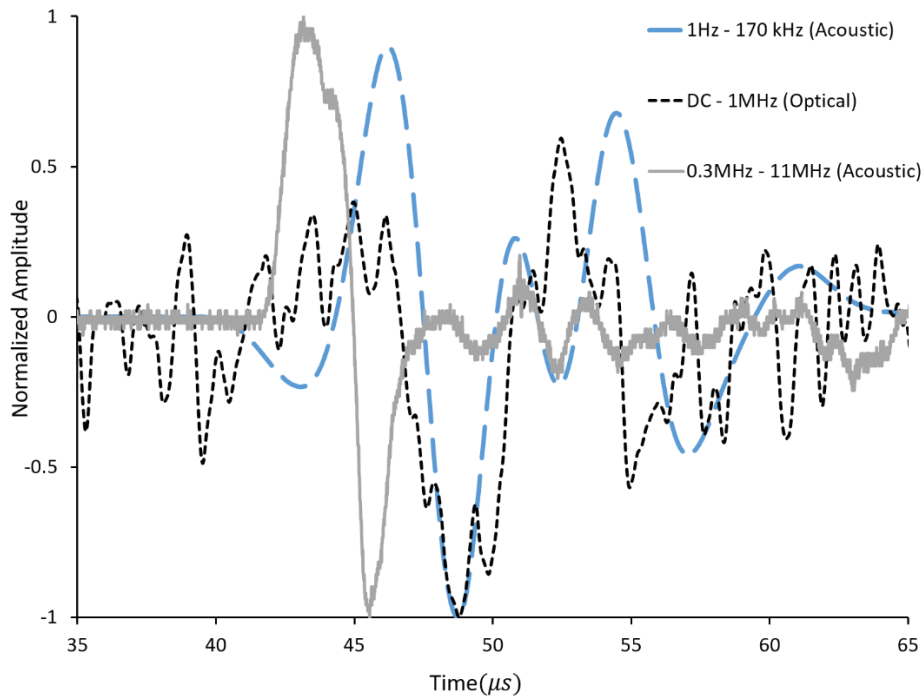

**Figure 7. Acoustic photonic coupling in laurdan containing vesicles made of 50:50 (w/w) mixture of DMPC and DPPC at 33℃.** *Normalized waveforms obtained using pressure sensors of different bandwidth compared to optical waveform (here shown up to 1MHz). This shows the temporal correlation between changes in pressure field and fluorescence emission exists at lower frequencies.*

As derived in Appendix B, if the energy of the dye molecule changes by $\Delta E_1$ by exchanging heat with the membrane, which changes its enthalpy by $\Delta H_2$ (Appendix B), then the entropy of a system near equilibrium $S \equiv S(E_1, H_2)$ (Appendix B) gives;

$$< \delta E_1 \delta H_2 > = kT_2{}^2 \left( \frac{\partial E_1}{\partial T_2} \right) \tag{3}$$

which can be further simplified using Maxwell's formula for radiation density giving;

$$< \delta\lambda_{max}\delta H_2 > = kT^2 \left( \frac{\partial\lambda_{max}}{\partial T} \right) \tag{4}$$

where $\frac{\partial\lambda_{max}}{\partial T}$ can be measured independently, plotted in fig. 8 shows a distinct peak at the phase transition in DPPC vesicles. Eq. (4) states that fluctuation in the maximum emission wavelength is coupled to the enthalpy fluctuations of the membrane environment, which in turn is given by $< (\delta H_2)^2 > = kC_{2_p}T^2$. The fluctuations are decoupled only when $\frac{\partial\lambda_{max}}{\partial T} = 0$. Furthermore, near phase transitions, it can be shown that (Appendix B);

$$C_p = \varepsilon\Delta\Gamma/T \tag{5}$$

Where $\Delta\Gamma = \sqrt{\Gamma^2 - {\Gamma_0}^2}$ is the *excess* FWHM of the emission spectra and can be measured experimentally. $C_p$ calculated using eq.(5) is plotted as $C_{local}$ in fig. 8 (inset) highlighting that it represents the local environment of the dye molecules. $C_{local}$ provides a good estimate for the macroscopic heat capacity $C_p$ measured in single phase region(35). Closer to the phase transition, while $C_{local}$ also has a maximum, the peak is not as sharp as those observed during macroscopic $C_p$ measurements. The sharpness of the peak in $C_p$ is known to represent the cooperativity of the phase transition(31). Here $\frac{\partial\lambda_{max}}{\partial T}$ is expected to depend on the cooperativity of the transition unlike $C_{local}$ calculated on the basis of spectrum width in Fig.8, but this requires further investigation. While Fig. 8 used data from a single study on DPPC vesicles, the correlation between the rate of shift of a spectrum as a function of state (pressure, temperature, pH, chemical potential), heat capacity and spectrum width has been observed in a variety of system consisting of artificial and even native membranes. Table 1 shows a list of such studies, indicating the correlations that were observed, including any correlation to biologically relevant functions, such as permeability and enzyme activity.

The maximum in $\frac{\Delta RP}{RP_0}$ near phase transition observed in the acoustic experiment (Figs. 3, 4 and 6) can now also be explained mechanistically. At phase transition, the heat capacity, compressibility, and acoustic absorption of the membrane have a maximum and therefore $\Delta H_2$ induced by the acoustics will be a maximum from which Eq. 4 indicates $\delta\lambda_{max}$ is maximum. Equation A.3 shows that $\frac{\Delta RP}{RP_0} \propto \delta\lambda_{max}/\Gamma^2$ and therefore the maximum in $\delta\lambda_{max}$ at phase transition will result in a maximum in $\frac{\Delta RP}{RP_0}$.

Note that in order to determine the heat capacity by Eq 5, it is necessary to measure $\Delta\Gamma$ which requires measuring the spectrum at least at three different wavelengths around the peak. The measurements would also allow for direct measurement of the dynamic changes in the heat capacity itself, i.e. second order effects of the acoustic impulse on the membrane.

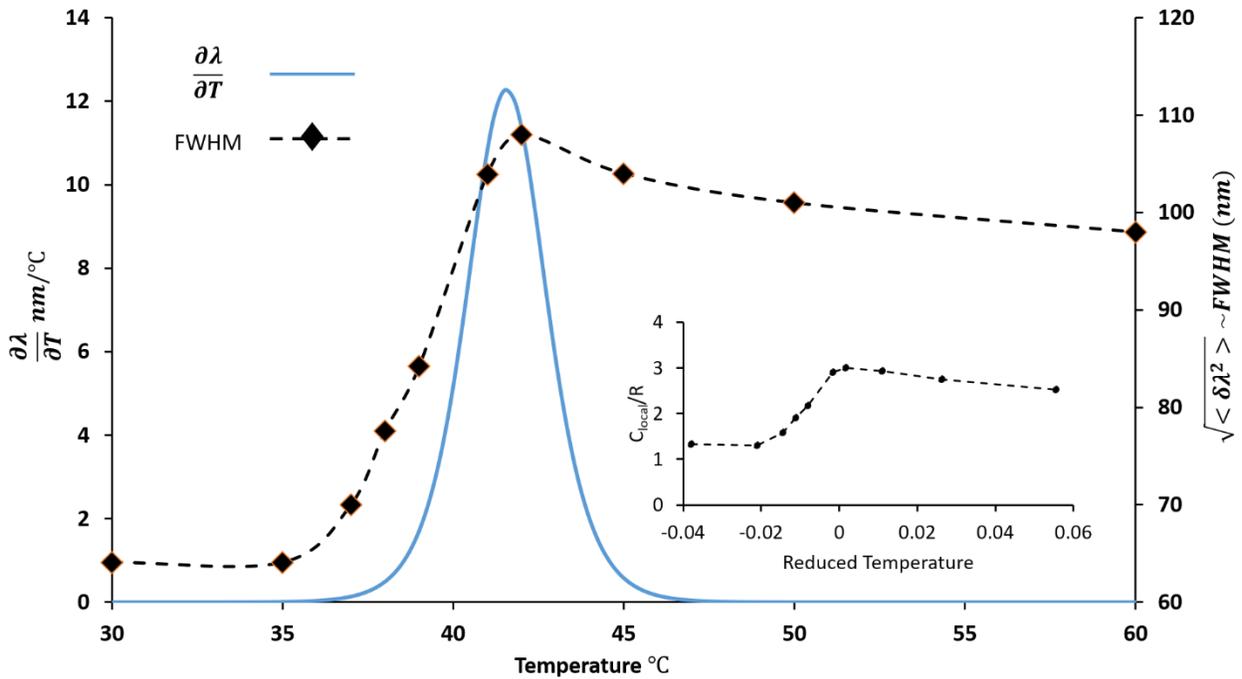

**Figure 8 Estimating heat capacity from the observed emission spectrum of Laurdan** *The rate of change of peak emission wavelength with respect to temperature $\frac{\partial\lambda}{\partial T}$ is plotted as a function of temperature and was obtained from a sigmoidal fit of previously reported peak wavelengths(see fig.2 in ref. (36)) for DPPC vesicles. From the same emission spectrum, FWHM was obtained and has been plotted on the secondary axis. The inset shows the specific heat calculated purely from the optical measurements of ref.35 and using a constant fit parameter of proportionality.*

Others have previously shown a direct correlation between heat capacity, compressibility and acoustic absorption of lipid membranes(2, 3, 33). Therefore, based on eq. (4) and (5) and the discussion above, the magnitude of the acoustic absorption can be, in principle, estimated from the width of the emission spectrum. This can prove useful in predicting acoustic absorption of a cellular or subcellular environment based on optical imaging (37).

| | System | Agent for state change | Properties measured | Observation | Reference |
|---|---|---|---|---|---|
| 1 | Laurdan in lipid vesicles | Temperature, pH and lipid composition | Emission spectrum and GP (T) | (1) FWHM(T)$_{max}$ in DPPC vesicles at 41C, coincides with PT of DPPC and [d(GP)/dT]$_{max}$. (2) [d(GP)/dT]$_{max}$ indicates phase transition in vesicles of different lipid composition and different pH | (36) |
| 2 | Laurdan in DMPG vesicles in low ionic strength | Temperature | Emission spectrum and GP (T) | (1) DPMG at low ionic strength has a broad PT starting at 17C and peaking at 35C. (2) FWHM(T)$_{max}$ around 35C which is in the PT region and near [d(GP)/dT]$_{max}$. | (12) |
| 3 | Laurdan in DPPC vesicles and Tamoxifen | Temperature, Tamoxifen concentration | Emission spectrum, GP(T), heat capacity C, Permeability | (1) Tamoxifen shifts DPPC PT and C$_{max}$ from 41C to lower temperatures. (2) A shift in C$_{max}$ due to tamoxifen coincides with the shift in [d(GP)/dT]$_{max}$ to 40C which is further consistent with the increase in FWHM. (3) No effect of tamoxifen on FWHM at 35 and 50C as expected from [d(GP)/dT]~0 (4) Temperature and tamoxifen concentration corresponding to FWHM also results in maximum membrane permeability. | (38) |
| 4 | Laurdan and Prodan in DSPC vesicles | Pressure and Temperature | Emission Spectrum | (1) FWHM for both Laurdan and Prodan, as a function of pressure, has a maximum at the pressure induced order-disorder transition at 63C. (2) Only the FWHM of Prodan has a maximum at pressure induced inter-digitated crystalline state at 49C while Laurdan is insensitive. | (13) |
| 5 | Prodan in DPPC | Cholesterol and Ethanol | Emission Spectrum | (1) FWHM as a function of cholesterol concentration has a maximum at 15% and 50C. Independently this coincides with the corresponding order-disorder transition in the DPPC/cholesterol phase diagram(3) | (23) |
| 6 | Laurdan in POPC and DPPC | Temperature and @-Tocopherol | Emission Spectrum | @-Tocopherol influences FWHM in POPC at 25C as [d(GP)/dT]#0 but not in DPPC as [d(GP)/dT]~0 at 25C | (39) |
| 7 | Laurdan in pig kidney basolateral membrane | Cholesterol | GP as a function of cholesterol concentration | Maximum N$^+$/K$^+$-ATPase activity coincides with cholesterol concentration corresponding to (d(GP)/d[cholesterol])$_{max}$ | (40) |
| 8 | Laurdan in lipid extracts from human skin stratum corneum (HSC) and bovine brain (BB) | Temperature, pH, lipid composition | GP (T), C(T) | C(T)$_{max}$ coincides with [d(GP)/dT]$_{max}$ for HSC extracts but not for BB extracts, again indicating that Laurdan might not be suitably located to detect all transitions. | (41) |
| 9 | Laurdan in plasma membrane from rat brain cortex | The concentration of Brij58, Triton, and Digitonin | Emission Spectra and G-protein activity | (1) Brij 58 and Triton induce a state of increased FWHM while Digitonin does not. (2) Brij58 and Triton concentrations corresponding to FWHM$_{max}$ induce a maximum increase in G protein activity while Digitonin does not. | (14) |

| 10 | Laurdan in HIV virus | Methyl-$\beta$-cyclodextrin and temperature | Emission spectrum | FWHM$_{max}$ coincides with [d(GP)/dT]$_{max}$ | (18) |
|----|----|----|----|----|----|
| 11 | Laurdan in brain phosphatidyl serine | Calcium concentration | Excitation and emission spectrum | Narrowing of the spectrum coincides with a blue shift on the addition of calcium | (42) |
| 12 | Laurdan in Escherichia Coli | Temperature and Chloramphenicol (cam) | Emission Spectra, steady-state and time-resolved | Peak in [d(GP)/dT] near physiological temperature which shifts to lower temperature on cam treatment. Narrowing of the emission spectra on cooling. | (43) |
| 13 | Laurdan in Rat liver microzomes | Temperature and fatty acids | Emission Spectra and GP | Significant narrowing and blue shift in spectra as a function of temperature | (44) |

**Table 1. Summarizing the experimental studies supporting the derived relationship between the fluorescence spectra of Laurdan and membrane susceptibility.** GP represents generalized polarization; C represents the heat capacity of the entire sample. GP(T) or C(T) for e.g. implies measured as a function of temperature. FWHM is the full-width half maximum of the emission spectrum.

Finally given the close correspondence between excited states during fluorescence and enzyme activity(45), the analysis can also be used to interpret the modulation of enzymatic activity and biological function in general by the lipid membrane and acoustic (mechanical) perturbations (mechano-sensing). For example, the width or the fluctuations of the emission spectrum of Laurdan has been shown to correlate with the simultaneously measured activity of several enzymes in the lipid membranes (Item 7 and 9, table 1) (14, 40). Thus acoustically induced changes in spectrum provide insights into possible mechanisms for acoustic modulation of enzyme activity based on this study. The mechanism can be explained in terms of top-down causality(46) starting from acoustically induced macroscopic changes in the thermodynamic state leading to microscopic changes in the fluctuations of the solvation shell of the embedded molecules. (a) The macroscopic changes in the thermodynamic state of the membrane are directly related to the acoustic radiation force. Near a phase transition, the heat capacity and the compressibility are maximum and therefore acoustic absorption is maximum (2) resulting in maximum acoustic radiation force and maximum state change for a given perturbation. (b) The macroscopic changes in the state also result in changes in heat capacity (second order effects of an acoustic impulse), which is related to changes in microscopic fluctuations of the solvation shell of the embedded molecule as discussed. It's the changes in the thermodynamic fluctuations that are responsible for changes in membrane permeability (1, 47) and enzymatic activity (27, 45, 48). Thus acoustic perturbation can effectively modulate transmembrane permeability and the enzymatic activity near a phase transition by modulating the state and hence the fluctuations of the solvation shell.

**Conclusions**

Lipid vesicles embedded with the solvation sensitive dye Laurdan were perturbed acoustically and the magnitude of the shift in the emission spectrum was shown to be related to the heat capacity of the system. Furthermore, it was shown that the acoustic response of a lipid membrane is dependent on the thermodynamic state and that the coupling is strongest in the vicinity of a phase transition of the lipids. These results imply that the efficiency of the membrane-based mechano-transduction depends on the thermodynamic susceptibilities of the collective membrane-protein system and also imply that acoustic waves are capable of switching the phase of the material which can result in changes in the activity of membrane-bound enzymes and the permeability of the membrane.

Due to the local nature of the fluorescence probes, it should now be possible to spatially resolve the heat capacity of cellular microenvironments from fluorescence imaging. Thus fluorescence measurements in living systems can also provide insights into the role of state during adaptation (49, 50), aging (51), and mechanotransduction (52). In particular, it can provide crucial insight into the role of thermodynamic state during native dynamic processes in biological membranes, such as action potentials. For example, in experiments done previously on nerve fibers during an action potential, the width of the emission spectrum was observed to shrink at the peak amplitude (8), which as per eq.(4) and (5) suggests a decrease in heat capacity and compressibility This is consistent with X-ray diffraction measurements, which suggested that the nerve membrane indeed gets stiffer at the peak amplitude (53). Finally, the measurement system described here could be used to determine the acoustic susceptibility of different cell types or a certain microenvironment, which in turn can help, for example, find better targets for acoustically triggered therapies.

**Acknowledgments**

S. Shrivastava and R. Cleveland are supported by Engineering and Physical Sciences Research Council (EPSRC) under Programme Grant EP/L024012/1 (OxCD3: Oxford Centre for Drug Delivery Devices). The authors thank James Fisk and David Salisbury for constructing the water tank with the shock wave setup. S. Shrivastava and M. Schneider thank Dr. Konrad Kaufmann (Göttingen), who introduced them to Einstein's approach to statistical physics and its importance for biological physics. His theory on the relationship between state and catalytic rate inspired this work. We would also like to thank Dr. Jeremy England and Dr. Christian Fillafer for their comments and suggestions.

## Appendix

### A. Optical measurements of spectral shift

We assume that the measured intensity from a dye embedded in the membrane can be expressed as $I(r,\lambda) \propto f(r)g(\lambda - \lambda_{max})$, i.e. the dependence on size (morphology) and wavelength can be separated into functions $f$ and $g$, respectively and r is the radius of the vesicles. The ratiometric parameter is defined as the ratio of intensity measured at two different wavelengths $\lambda_1$ and $\lambda_2$ is defined as

$$RP = g(\lambda_1 - \lambda_{max})/\, g(\lambda_2 - \lambda_{max}) \tag{A.1}$$

We note that the ratiometric parameter $RP$ is related to the generalized polarization $GP = \frac{RP-1}{RP+1}$. Both RP and GP have been widely used to study lipid phase transitions {REF}. A perturbation of the spectrum in terms of $RP$ is given by;

$$\frac{\Delta RP}{RP} = -\Delta\lambda_{max}\left(\frac{g'(\lambda_1-\lambda_{max})}{g(\lambda_1-\lambda_{max})} - \frac{g'(\lambda_2-\lambda_{max})}{g(\lambda_2-\lambda_{max})}\right) \tag{A.2}$$

Choosing $\lambda_1$ and $\lambda_2$ such that $\lambda_1 < \lambda_{max} < \lambda_2$ simplifies the interpretation, as then the term within the bracket, that quantifies the shape of the spectrum, is always positive. In the experimental section, we use $\frac{\Delta RP}{RP}$ to quantify the perturbation of the emission spectra due to an acoustic field.

In the case of $g$ having a Gaussian form, i.e. $g(\lambda - \lambda_{max}) \propto e^{-\left(\frac{\lambda-\lambda_{max}}{\Gamma}\right)^2}$, then Eq A.2 becomes:

$$\Delta RP/_{RP} = \frac{\Delta\lambda_{max}}{\Gamma^2}2(\lambda_2 - \lambda_1) \tag{A.3}$$

Thus $\frac{\Delta RP}{RP}$ is proportional to the spectrum shift and inversely proportional to the spectral width squared.

## B. The relation between the emission spectrum of the dye and the heat capacity of the membrane

Fig.S1 presents a simple outline of the fluorescence phenomenon considered here from the dye's perspective.

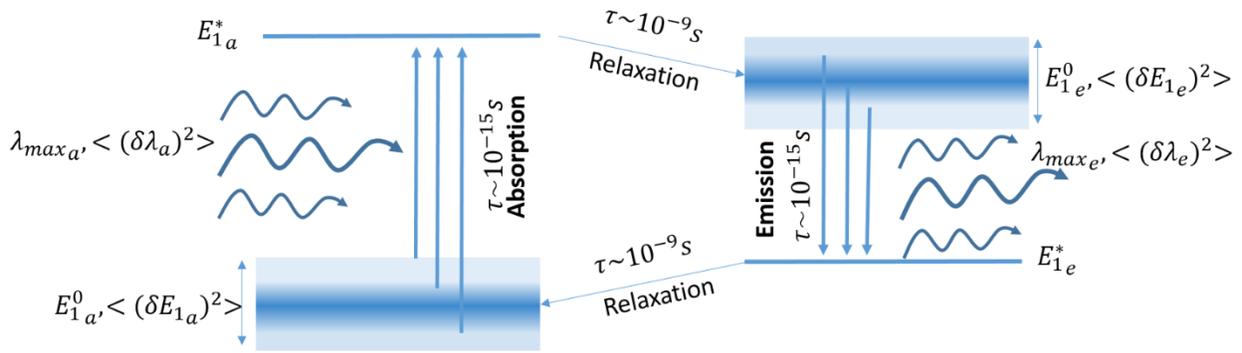

**Figure S1 Approximate representation of fluorescence absorption and emission.** *The schematic outlines how the variations in the photon wavelengths $<(\delta\lambda)^2>$ during absorption or emission process are related to the variations $<(\delta E_1)^2>$ in the ground state energy $E_1^0$ of the fluorophore. For absorption, the ground state dipole of the fluorophore is in equilibrium with the environment, while for emission the ground state transition dipole is assumed to have reached an equilibrium with its solvation shell (10). The excited states $E_1^*$ immediately upon absorption or emission is assumed to be independent of solvent relaxation phenomenon (see text). The subscript 'a' and 'e' for absorption and emission respectively, have been dropped in the main text for the sake of simplicity.*

The dye (subsystem 1) interacts with the rest of the interface (subsystem 2 consisting of lipids and water). Let $E_1$ be the internal energy of the dye and $H_2$ be the enthalpy of the rest of the interface. The number of ways the total energy can be distributed in these two compartments is a function of the entropy of the entire system, given that the two systems interact and the individual probabilities are not independent, i.e. $(W(E_1, H_2) \neq W(E_1).W(H_2))$ (54). The probability distribution function is related to the entropy of the interfacial system and is defined as $S(E_1, E_2)$ (55, 56) via the Boltzmann principle;

$$dW = K.e^{S(E_1,H_2)}dE_1 dH_2 \tag{B.1}$$

where $K$ is a constant. Realizing that the most probable energy state $E_1^0$ of the dye lies at the maximum of entropy potential, the Taylor expansion of the entropy up to the second term near $(E_1^0, H_2^0)$ can be written as;

$$\Delta S(E_1, H_2)_{E_1^0, H_2^0} = \frac{\partial S}{\partial E_1} \delta E_1 + \frac{\partial S}{\partial H_2} \delta H_2 + \frac{1}{2} \frac{\partial^2 S}{\partial E_1^2} (\delta E_1)^2 + \frac{1}{2} \frac{\partial^2 S}{\partial H_2^2} (\delta H_2)^2 + \frac{\partial^2 S}{\partial E_1 \partial H_2} \delta E_1 \delta H_2 \quad \text{(B.2)}$$

where the derivatives are taken at $E_1 = E_1^0$, $\delta E_1 = E_1 - E_1^0$ and $H_2 = H_2^0$, $\delta H_2 = H_2 - H_2^0$. As there is an entropy maximum at $(E_1^0, H_2^0)$, we have $\frac{\partial S}{\partial E_1} = \frac{\partial S}{\partial H_2} = 0$ and using the definition of temperature $\left(\frac{\partial S}{\partial E}\right)_v = \frac{1}{T}, \left(\frac{\partial S}{\partial H}\right)_p = \frac{1}{T}$ and heat capacity $C_v = \frac{\partial E}{\partial T}$ $C_p = \frac{\partial H}{\partial T}$, eq.(B.2) can be reduced to (57);

$$\Delta S(E_1, H_2)_{E_1^0, H_2^0} = -\left\{ \frac{1}{2} \frac{(\delta E_1)^2}{C_1 T_1^2} + \frac{1}{2} \frac{(\delta H_2)^2}{C_2 T_2^2} + \frac{\delta E_1 \delta H_2}{T_2^2} \frac{\partial T_2}{\partial E_1} \right\} \quad \text{(B.3)}$$

where $\delta E_1 = E_1 - E_1^0$, $\delta H_2 = H_2 - H_2^0$ and $T_1$ and $T_2$ represent the microscopic temperatures of the dye subsystem and environment, respectively. Note that in this appendix $T$ has the units of energy, i.e. normalized by Boltzmann constant. The microscope temperatures only acquire physical significance from the macroscopic measurements of temperature $T$ when the system is assumed to be in local equilibrium and hence $T_1 = T_2 = T$. Substituting eq. (B.3) into eq. (B.1) and comparing the coefficients with a Gaussian distribution results in the following two relations

$$< (\delta H_2)^2 > \sim C_p T^2 \quad , \quad \text{(B.4a)}$$

and

$$< \delta E_1 \delta H_2 > \sim T^2 \left( \frac{\partial T_2}{\partial E_1} \right)^{-1} \quad \text{(B.4b)}$$

Going back to the fluorophore membrane system, the internal energy $E_1$ of the dye (subsystem 1), is directly related to the mean radiation energy $E_r$ absorbed or emitted by the fluorophore as per the first law;

$$|E_1^* - E_1| = E_r(\lambda) \quad \text{(B.5)}$$

Here $E_1^*$ is the energy of the electronic state immediately upon absorption or emission which has a scale of $\sim 10^{-15} \text{sec}$ (58) and can be assumed to be independent of any rearrangement in hydration shell ($\sim 10^{-9}$sec) (10, 59). Therefore any variation in $E_r(\lambda)$ due to rearrangement of the hydration shell can be attributed mainly to variations in $E_1$, i.e. only the relaxed state energy distribution of the dye molecules. The photon intensity has been assumed to be low so that only a small fraction of dye molecules is excited at a time and the pressure and temperature are not changed significantly(60, 61). From Maxwell theory, the spectrum $\rho_\lambda$ (radiation density) and the corresponding mean radiation energy absorbed or emitted $E_r(\lambda)$ are related as (62)

$$E_r(\lambda) = \frac{L^3 \lambda^2}{8\pi c^2} \rho_\lambda \qquad (B.6)$$

For the emission process, if $\lambda_{max}$ is the observable that determines the state phenomenologically based on Einstein's description of the entropy potential (63), then differentiating eq. B.6 and substituting in eq, B.5, also using $\left(\frac{d\rho_\lambda}{d\lambda}\right)_{\lambda_{max}} = 0$ and for a normalized distribution ($\rho_{\lambda_{max}} = 1$), we have $\delta E_1 \sim \delta E_r \sim (\lambda_{max} \delta \lambda_{max})$. Now by propagating the fluctuations in $E_r(\lambda)$ to $\lambda$, and using eq.(B.4), (B.5), (B.6) following relation can be derived after dropping the constants

$$< \delta \lambda_{max} \delta H_2 > \sim T^2 \left(\frac{\partial T_2}{\partial \lambda_{max}}\right)^{-1} \qquad (B.7)$$

This shows that $\delta \lambda_{max}$, i.e. the fluctuation or perturbation in the wavelength of radiation absorbed or emitted by the ensemble of dye molecules (subsystem 1), is coupled to the fluctuations or perturbation $\delta H_2$ of the membrane via the state dependent coupling parameter $\left(\frac{\partial T}{\partial \lambda_{max}}\right)^{-1}$.

While the derivation so far is applicable to all states, assumptions can be made in the vicinity of a phase transition that allows a direct relation between the spectral width and heat capacity. The coupling term is plotted as a function of temperature in fig. 8 (Main text) for DPPC MLVs and shows a strong correlation with the heat capacity $C_p$ (64) allowing us to write as an empirical result, with $\gamma$ as an empirical constant;

$$\left(\frac{\partial T}{\partial \lambda}\right)^{-1} = \gamma \ C_p = \frac{\partial H}{\partial T} \qquad (B.8a)$$

$$\Delta \lambda_{max} = \gamma \Delta H_2 \qquad (B.8b)$$

Thus fluctuations in $\lambda_{max}$ are directly correlated to the energy of the surrounding membrane and the strength of coupling is determined phenomenologically by the constants of proportionality $\gamma$ in eq.(B.8). The correlation derived here is valid near the transition region, but its general applicability has been discussed previously in detail by others for fluctuation correlations in lipid systems (31, 65, 66).

Assuming a correlation coefficient of $\frac{<\delta\lambda_{max}\delta H_2>}{\sqrt{(\delta\lambda_{max})^2}\sqrt{(\delta H_2)^2}} = 1$ and using eq. (B.7) and (B.8) one can write;

$$C(T) = \varepsilon\Delta\Gamma/\text{T} \tag{B.9}$$

Where $\Delta\Gamma = \sqrt{\Gamma^2 - {\Gamma_0}^2}$ is the *excess* FWHM from spectrums of Laurdan, which has been plotted for DPPC in Fig.8 using previous data (36), $\varepsilon$ is a phenomenological constant of proportionality and $\Gamma_0$ is the spectral width of the dye in vacuum corresponding to its intrinsic specific heat at room temperature. To distinguish the heat capacity of the entire sample as measured by a calorimeter and the heat capacity calculated using eq. (B.9), which represents the local environment of the dye, we represent the later by $C_{local}$. The inset in Figure 8 shows $C_{local}$ as obtained from Laurdan as a function of T*, using $\varepsilon = 10R.\,nm^{-1}$, where R is the gas constant and $\Gamma_0 \approx 50nm$ approximated as FWHM of Laurdan in n-hexane (67). The estimated specific heat profile is remarkably similar to quantum statistical $C_p$ calculated previously (35).